# Emerging Reflections from the People of Color (POC) at PERC Discussion Space

**List of Participants:** Carolina Alvarado, Deepa N. Chari, Ximena C. Cid, Geraldine L. Cochran, Joel C. Corbo, Ayush Gupta, Simone A. Hyater-Adams, Raina Khatri, Alexis V. Knaub, Yuhfen Lin, Mike A. Lopez, Camila Monsalve, Gina M. Quan, Laura Ríos, Chandralekha Singh, Brian Zamarripa Roman

If you have any questions, comments, or concerns about this document, or interests in joining the email list you can contact us via email at pocinper@gmail.com.

## Introduction

### About the POC in PER Discussion Space

In 2017, Geraldine Cochran and Alexis Knaub hosted a PERC parallel session called the POC in PER Discussion Space. POC in PER stands for People of Color[1] in Physics Education Research. Both were becoming more aware of challenges that POC face in PER at all levels of the academy. The goals were to provide a space where POC could share their experiences and validate the experiences of others.

Prior to hosting the space, Cochran and Knaub talked about the goals of the space and rules for creating an environment in PER for POC. They sought to create an environment for POC where topics emerged based on what attendees were willing to share, rather than choose the discussion topics. One of the outcomes of the space was to write a paper that informs the broader PER community of the discussion and to offer suggestions to make PER more inclusive and equitable environment for POC.

Attendees shared various racist incidents that occurred at the conference as well as outside of the conference, at venues in the location or at home in their groups. Due to the feedback from the 2017 session, Cochran and Knaub hosted a similar session at PERC 2018. In addition to the formal session, an after-care session was available for participants that was facilitated by a representative from Lumos Transforms, a company that focuses on reducing stress and healing trauma. This was funded by a PERTG grant. Brian Zamarripa Roman and Simone Hyater-Adams, with support from Alexis Knaub and Geraldine Cochran, will host another POC in PER Discussion session at PERC 2019. Due to the emergent needs and interests of the participants of the past sessions, the group continues to engage in other supportive activities that are not described in this paper.

### About this Paper

This paper reflects both the discussions from 2017 and 2018. Attendees volunteered to lead sections that summarized the broad themes from discussion. As this is intended to be a collectively-created document that reflects the discussion of the POC in PER who attended at least one of these sessions, other attendees have helped edit this document.

---

[1] In this paper we choose to capitalize the term People of Color to center and highlight the agency of people who experience similar forms of marginalization due to their non-white racial, ethnic, and national identities in the American context. We align this practice with those of other authors in search of social justice and equality.
https://radicalcopyeditor.com/2016/09/21/black-with-a-capital-b/

We emphasize that this discussion paper reflects the experiences of the attendees who contributed to discussion in the sessions and may not reflect the experiences of all People of Color in the PER field.

# Professional Concerns
*Section led by Alexis V. Knaub and Simone Hyater-Adams*

In the session, an emergent topic from the discussion was the collective experience of POC in PER not getting the same professional opportunities as their white colleagues. This comparison was emphasized because it not only highlights the disparities but also points to another issue: that POC in PER, despite facing challenges in receiving opportunities, are evaluated and assessed in the same arena and based on the same criteria as white counterparts who have not faced the same challenges. For researchers in any discipline, professional opportunities are not only needed for intellectual growth but also to succeed in the field. Below we list the patterns of experiences that were brought up in the session about challenges with professional opportunities.

**Participants noted times where they are not acknowledged as experts in areas where they are experts**
Participants have observed that despite being accomplished and knowledgeable in a research area, they are not treated as such in comparison to white individuals. One example is being an invited speaker. In the session, participants discussed observing that individuals invited to present in PERC keynotes are rarely people of color. Another example is when white individuals are more frequently sought for their knowledge when a POC has the same knowledge. This occurs even when the white individual has less experience and knowledge than the individual of color.

Additionally, some participants describe a double standard where colleagues assume they are experts in equity, diversity, and inclusion simply because of their identity, yet those who are experts are less valued for this expertise than their white colleagues. Participants often care more about equity efforts because of experiences of oppression that impact us in our day to day lives professionally and personally.

**Participants noted incidents where they are denied opportunities to collaborate**
Participants in the session discussed the pattern of white researchers being less likely to suggest a POC as a collaborator. In some instances, they simply are overlooked. More egregiously, participants have noticed that white researchers have dismissed POC as collaborators for reasons that also applied to other white collaborators on the same project. Succinctly, the white researchers in charge held a double standard. In other cases, when POC were recommended as collaborators white researchers have indicated a preference for working with individuals with whom they already had a working relationship, which were all white researchers. If this had been one incident, it may have seemed to be an individual bias; however, our collective discussion confirmed that this is a common occurrence.

**Participants noted that they do not receive the same opportunities to develop their professional networks**
Professional networks are important for many reasons including forming collaborations and finding job opportunities. Even if initial contact does not result in a concrete benefit, they often do in the future (see: the "old boys' network."). However, some participants have noted that they do not receive the same opportunities to develop their professional networks. This includes being introduced to key contacts. These opportunities have been presented to white counterparts. We note that participants have stated that this is not just a matter of getting POC to ask for these opportunities. They have observed that white counterparts are sometimes just handed these opportunities to develop their networks. Sometimes, participants have asked for support in developing their networks and have been denied this, even by their PIs.

**Participants noted that there are often expectations of them to take on additional work that is professionally undervalued**
Service work is part of professional life, but due to their identity—and often intersecting identities—POC may take on more work than white counterparts. This can be for many reasons, including the few number of POC in academia who can serve in these roles. Some of this work includes mentoring students from underrepresented and marginalized groups, providing emotional support for students and colleagues, and serving on committees related to diversity. This can take up considerable time as well as emotional energy. While many participants in this session are willing to do this work and feel that it is often an obligation, they also acknowledge that this is simultaneously expected of them and also not recognized as a meaningful professional contribution. Some participants have noted that when white counterparts do some of this work, they are lauded and professionally awarded.

For some, working for and supporting their communities through service work is not seen as work but a means of survival as it is ingrained in their cultural upbringing. Understanding how to frame this work as "work" as opposed to creating and maintaining connections to their home communities/identities is something that is not discussed and can be a challenge in terms of balancing work obligations. To frame this type of service as work that should be compensated in some way can create a conflict as work often needs to be quantified with reports, time management, and tenure applications and this type of service is rooted in cultural connections.

**Participants noted incidents where they have been disparaged by colleagues**
No one is above critique, regardless of identity. However, several POC have noted that individuals within PER have attacked some participants' professional reputation behind their backs and warned others not to work with them. While "whisper networks"[2] can be useful to protect individuals from harm when no viable alternative exists, such as harassment, there appears to be no legitimate reason that such warnings were needed against these individuals. These individuals did not appear to address whatever grievances they had with the POC, and instead, chose to warn others about the POC or in some cases to reach out to people completely unrelated to the issue/situation.

# Dealing with Racism in General Society
*Section led by Brian Zamarripa Roman*

As a society we are aware of instances of police brutality and everyday discrimination towards People of Color; however, as POC we deal with the reality that this may happen and has happened to us. This discrimination continues, but increased access to technology and the current political climate in the

---

[2] A "whisper network" involves discreetly informing others that an individual is harmful. While we acknowledge that whisper networks can be useful, we also see them as more of a bandage rather than a good solution. They operate on the assumption everyone will receive the information, ask that the potential victim protects themselves, and do not actually handle the problematic entity.

United States have made it more prominent in the public eye [1]. The political climate and the current U.S. administration has allowed racism to be more overt and deadly. Although, we may not experience blatant racism within the AAPT community as often as in the general public, it is expected once we step outside of the hotel/conference center and interact with local people, businesses, and authorities. In this section, we have a more detailed discussion on the instances of racism and discrimination that may be experienced in spaces outside conference settings and bring to light experiences of discrimination of POC during conferences which were discussed in the POC space.

The 45th President of the United States has normalized rhetoric that dehumanizes POC and perpetuates white supremacy. From his comments about Mexican immigrants "bringing drugs… crime" and claiming they are rapists, referring to African and Caribbean countries as shitholes, and setting up a Muslim travel ban, to his marking of neo-Nazi groups as "fine people" [2], his language has afforded undercover racists the justifications to act on their biases. As we organize AAPT conferences and set locations, we must not only recall that we are operating within a country led by a man with these ideologies, but more importantly that we are operating within a country where nearly half the voting population supported and elected this man as its leader. In other words, his leadership is a product of an established system that has existed before him and is likely to remain afterwards [3]. POC attending AAPT conferences must interact with the country in multiple capacities from businesses, such as transportation services, airports, lodging, restaurants; other people cohabitating these spaces; and law enforcement and authorities, with the risk of a racist confrontation at every stage. Inexcusably, these have happened, as discussed below.

When it comes to businesses, employers and employees are not exempt from racist actions. This can start off as early as trying to get an Uber ride and waiting longer or having rides cancelled [4]. At airports, POC often get unfair treatment by airport security and are kicked out of planes for minor complaints [5-8]. Once on-site, conference goers will need to go to restaurants, where they may be denied services that are provided to white people in a similar situation [9].

POC will not only interact with employees, but other customers as well. People in society at large have confronted POC speaking languages other than English with the claim that English is the language of the United States, which have led to the involvement of law enforcement [10,11]. There are also conference goers that will stay at places at a distance from the conference such as Airbnbs, which can be compromised by racism as POC have had authorities called on them for not being recognized in neighborhoods [12]. Recently, there has been a surge of stories of POC having authorities called on them for minor reasons [13]. An example of this is the CVS employee that called the police on a Black woman because he did not believe she had a valid coupon or the two Black men that were arrested for waiting at a Starbucks [14,15]. These two scenarios could easily happen to POC AAPT attendees since conference attendees often need to go to pharmacies or coffee shops.

The moment authorities are involved is when things may take a deadly turn. Over the last few years we have been hyper aware of police brutality inflicted on POC. Hundreds of POC have been killed at the hands of local authorities for minor infractions. These outcomes between POC and law enforcement have changed from "can happen" to "likely will happen" in terms of mental preparation when traveling to new spaces for conferences and other academic programs. Federal authorities also pose a threat since conferences are hosted in states such as Texas (SB 4) and Arizona (SB 1070) that enforce laws preventing interference with agencies such as U.S. Immigration and

Customs Enforcement (ICE). This means that ICE can interrogate any person they want and not even local and state authorities can interfere.

These incidents occur in general society and new incidents occur daily; thus, they are not only a possibility, but a reality during conferences. POC in the discussion space have been detained, handcuffed, and had their citizenship questioned by local law enforcement when walking back to their Airbnb. Others have also been questioned by hotel staff about their guest status during conferences on various occasions and conference locations. Individuals in the space have also been pulled over by police in their Uber on their way to restaurants at night and accused of drunk driving due to the driver having a non-English accent. A large group of POC were kicked out of a bar after a minimal altercation with one of the individuals, while the white perpetrator was allowed to remain in the premises. Additionally, the AAPT conference in Texas raised serious concerns regarding the possibility of POC being interrogated and detained. These incidents were the ones willingly shared by POC in the space, however it is likely other traumatic racist experiences have occurred that were not brought to light. These experiences, as well as those happening in general society, lead to POC in PER becoming hyper aware of potential racist incidents and carrying a burden that their white counterparts may not be aware of.

AAPT is an organization that promotes and encourages diversity and the Physics Education Research Community should endeavor to support diversity. As a starting point, the community should be aware of the triad of agents (businesses, individuals, and authorities) that potentially perpetuate racism and the ways they interact with POC who are AAPT attendees. We may not be able to end racism, but it is necessary to take steps so that POC invested in our organization face minimal risks of racist confrontations. One concrete step that AAPT and PERC can take is considering the impacts of choosing conference sites in states that have policies and laws that further the oppression of certain groups (e.g., laws against gender neutral bathrooms, laws supporting ICE) and developing support structures that mitigate negative impacts.

# Reflections on the Physics Allies

*Section led by Ayush Gupta and Simone Hyater-Adams*

**Positionality for Ayush**
I was approached by the main coordinating team of Physics Allies in 2016 and 2017 to serve as a regular-member of the organization and accepted. I also attended the 2017 PERC POC Session where concerns about Physics Allies were raised. I find myself at an intersection, having membership in both groups and wanting to do justice to multiple perspectives. That can be perceived as a weakness. It can also have benefits—membership within Allies and personal relationships with specific members of the core leadership team allowed me to understand their motivations, but also push hard to make heard the concerns of community members of color. In writing this summary, I also am aware of and acknowledge my personal capital within the PER community, and the ways in which I stand at complex intersections, participating simultaneously in oppression, resistance, and alternative story-telling.

[Physics Allies](#) was formed as a group within AAPT in response to concerns from within our community towards unsafe practices in our community: incidents of racialized/gendered microaggressions, alcohol-fueled social events, etc. Drawing on the model of [Astronomy Allies](#), a small group of people came together to create a sub-community of people who can provide support to members of the physics education research community in moments of such incidents and advocate for and create opportunities for safer social spaces during our conference meetings. (For more information, see the July 2018 [PERTG newsletter](#) that provides information on the group's activities, scope, and future plans.)

**Reflections from the Session**
In the POC in PERC 2017 session, several participants brought up the Allies group. There were many who had negative experiences with members in the group individually and with the service that the group provides. There was a critique of the group's name, as the term "Allies" is commonly misconstrued as a title one can hold as an identity. In reality, allyship is an action, and holds no meaning unless it is describing the act of supporting and uplifting a group that is marginalized. At the session, participants shared a variety of concerns:
- Experiences of microaggressions from a person wearing an Allies button
- The complicated issue of seeking the help of an Ally, especially when they occupy positions of power within our community: how do we disentangle that from future professional interactions and/or access to professional opportunities?
- The use of the tag, "Allies"
- How allyship can, at times, become a vehicle for gaining community capital
- That the Physics Allies group was primarily white (which raises the question of whether the 'organizational gaze' of Physics Allies centers the perspectives of white folks towards issues)
- Whether and how folks who are serving in the role of Allies are engaging in self-education
- Lack of clarity around the purpose and scope of the organization

Some of the concerns were born out of direct experiences of certain participants; others from concerns about allyship as a mode of action; yet others hearkened to gaps in our understanding of an organization that claims Ally status. At that session, Ayush committed to raising some of these concerns with the core leadership of Allies.

Using the [web-based feedback form](#), he sent a detailed statement of concerns, ways to understand the concerns, and specific actions Physics Allies could take in response to each concern. That journey was a bit rough, logistically and emotionally. Responses to the statement varied from deeply understanding the concerns to framing them as a "messaging" problem. The responses highlighted to me two things: (i) the reality of fragile, defensive responses from some individuals in the face of concerns and the difficulty of calling them out on that fragility and defensiveness and (ii) the diversity of responses, in that there are also individuals who want to sit and reflect deeply on concerns raised, learn from them, and do the hard work of organizational change. It is important to tell this story without essentializing either point.

**Updates on Physics Allies**
The [PERTG section on Physics Allies](#) acknowledged and responded to some of these concerns. We are quoting that here so it can become available to us for reflection and discussion: "Our current activities are only small steps toward our vision of a more safe and welcoming community. The Physics Allies will need to continually refine its structure and function based on community feedback. We are aware that it is our responsibility to hear and work with, rather than for, those oppressed. As allies we are not all equally well-prepared to deal with every form of oppression, which is complicated even more by intersectionality. For these and other reasons, the term "Allies" is problematic for some in the PER community. We also acknowledge that we will make mistakes as Allies and, at times, ourselves act in ways that may offend, cause harm or perpetuate systems of oppression. We hope that when these moments are brought to our attention that we can turn toward them as learning opportunities."

Physics Allies also listed some action items:
- Increase the representation of different excluded identities within our organization and leadership.
- Engage in critical reflective discussions, self-education, and/or trainings.
- Aim to be transparent about the preparation that individual physics allies have pursued.
- Brainstorm about grassroots events that could raise community awareness of issues of safety and oppression within our community.

Ayush was asked to join the main coordinating team of Allies, and he accepted. In part, Ayush felt that since he had raised concerns to the group, he should also play a role in addressing them going forward. In part, he thinks that the overall mission of Physics Allies makes sense to contribute to, though supporting that mission responsibly would remain an uneven landscape. Most likely, the articulated action items listed above aren't sufficient. Some uncertainties also remain around specific ways in which some action items can be achieved. But, perhaps, finding our footing on that uneven landscape is needed work.

**Final Update:** As of now, the Physics Allies group has been dissolved.

# Discussions about POC participants in PER
*Section led by Geraldine L. Cochran*

POC in PER are excited about the broadening of participation in physics and the increased awareness to engage in PER that includes student participants from underrepresented socio-demographic groups[3]. Some of the participants have served as participants in PER studies and/or often relate closely with the students that are being described in physics classrooms and the student participants. Thus, it is very hurtful to hear students being referred to using disparaging terms. Most notably, students have been labeled or categorized as "weaker", "lowest", "low-level", "low-income", "not as good", and other such terms that we have heard in presentations and seen printed in publications. The use of terms like these demonstrates a lack of understanding of the inequalities in our society and more egregiously it attributes a student's circumstances and performance, which is based on a variety of factors, to their identity. Or put another way, these terms create an unofficial "norm" population that it is assumed that everyone has equal access to without acknowledging the systemic inequalities that have specifically and intentionally been put into place to advantage some populations over other populations, in this case access to quality education/preparation.

We suggest asking the following questions:
- Would I use this label or categorization for my students if they were present in the audience? For example, would I refer to a student as low-level in their presence?
- Have I attributed a circumstance or situation beyond a student's control to their identity?
- What is the purpose of this category/quote/information? What does the audience gain by including it? What do they lose if I do not include it?
- Does my research perpetuate harmful stereotypes and narratives about student success or the success of other populations?
- Would I be comfortable if someone used this kind of terminology when referring to me?

Participants have also noticed that some of the qualitative research in PER relies on shock-value and claims of -isms (including racism, sexism, etc.). We see presentations that bring out the worst of what people say about students from marginalized groups in a sort of trauma-porn fashion without a focus on how to dismantle the structures that lead people in positions of power to make such claims. When presenting qualitative research, what are our goals in sharing participant responses that include blatantly racist, sexist, or homophobic remarks? People from marginalized groups are well aware that some individuals hold hatred toward them and deeply held beliefs that they are inferior. Thus, we find the repeated presentation of such information unproductive and troubling. How does this move our field further and how does it support those from marginalized groups who are impacted not only by the words, but likely by the actions of these individuals as well? In addition to finding it emotionally distressing to sit through such presentations, we find it hurtful when our colleagues laugh at such statements. We understand that nervous laughter is a natural human response for some people when they are caught off guard and in shock. However, we would like the community to be aware of the feelings of their colleagues in the audience who do not find these anecdotes humorous, but who instead are emotionally unsettled and often triggered by such comments. We ask that researchers in PER think deeply about the aims and goals of their research presentations and how the presentation of their results might impact a diverse audience. Although this section has focused on research with student populations, the aforementioned pertains to all research populations, including students at all levels, postdocs, faculty and other professionals, and members of local communities.

---

[3] Kanim, S. & Cid, X.C. (2017) The demographics of physics education research. *In review.* https://arxiv.org/abs/1710.02598

Finally, we ask that our colleagues engaged in PER consider equity in the design of their research.

## Concluding Questions

The themes discussed in the session all align with patterns of systemic and structural racism. These issues hinder POC in PER's participation in the conferences and in PER at large. If you are reading this and wondering what you can do to change the experiences that POC have in the PER field, consider the following questions:

1. What are you doing to educate yourself and your white peers about anti-racist practices?
2. How might your work perpetuate some of the ideas and patterns of experience discussed here?
3. How might your classroom, department, or institution's policies be perpetuation some of the ideas and patterns of experiences discussed here?
4. In what ways can you set up accountability measures for yourself so that you can receive feedback and change your actions?